\documentclass[useAMS,usenatbib,useAMS]{mn2e}

\usepackage{times}

\usepackage{graphicx}
\usepackage{amsmath, amssymb}
\usepackage{txfonts}
\usepackage{amssymb}

\usepackage{graphicx,graphics}
\usepackage{natbib}
\usepackage{txfonts}
\usepackage{amssymb}

\usepackage{graphicx,graphics}
\def\km{\,{\rm km}}
\def\cm{\,\rm cm}
\def\s{\,{\rm s}}

\def\msun{\,{\rm M_{\odot}}}
\def\erg{\,{\rm erg}}
\def\kev{\,{\rm keV}}

\def\yr{\,{\rm  yr}}
\def\kpc{\,{\rm kpc}}
\def\Gpc{\,{\rm Gpc}}
\def\Gyr{\,{\rm Gyr}}
\title[SGRB and GW from eccentric compact binaries]{Short Gamma-Ray Bursts and Gravitational-Wave Observations from Eccentric Compact Binaries}
\author[Tan, Fan, Wang]{Wei-Wei Tan$^{1,2,3}$, Xi-Long Fan$^{1}$ \& F. Y. Wang$^{2,3}$\thanks{E-mail:
fayinwang@nju.edu.cn}\\
$^1$Hubei University of Education, 430205, Wuhan, Hubei, China\\
$^2$School of Astronomy and Space Science, Nanjing University,
Nanjing 210093, China\\
$^3$Key Laboratory of Modern Astronomy and Astrophysics (Nanjing
University), Ministry of Education, Nanjing 210093, China}
\begin{document}
%\thanks{E-mail:wwtan@nju.edu.cn}
\date{}

\pagerange{\pageref{firstpage}--\pageref{lastpage}} \pubyear{}

\maketitle

\label{firstpage}

\begin{abstract}

Mergers of compact binaries, such as binary neutron stars
(BNSs), neutron star-black hole binaries (NSBHs), and binary black
holes (BBHs), are expected to be the best candidates for the sources
of gravitational waves (GWs) and the leading theoretical models for
short gamma-ray bursts (SGRBs). Based on the observations of SGRBs,
we could derive the merger rates of these compact binaries, and
study the stochastic GW backgrounds (SGWBs) or the co-detection
rates of GWs associate with SGRBs (GW-SGRBs). But before that, the
most important thing is to derive the GW spectrum from a single GW
source. Usually, GW spectrum from a circular orbit binary is
assumed. However, observations of the large spatial offsets of SGRBs
from their host galaxies imply that SGRB progenitors may be formed
by the dynamical processes, and will merge with residual
eccentricities ($e_{\rm r}$). The orbital eccentricity has important
effect on GW spectra, and therefore on the SGWB and GW-SGRB
co-detection rate. Our results show that the power spectra of the
SGWBs from eccentric compact binaries are greatly suppressed at low
frequencies (e.g., $f\lesssim1~\rm Hz$). Especially, SGWBs from
binaries with high residual eccentricities (e.g., $e_{\rm
r}\gtrsim0.1$ for BNSs) will hard to be detected (above the
detection frequency of $\sim100~\rm Hz$). For the co-detection rates
of GW-SGRB events, they could be $\sim1.4$ times higher than the
circular case within some particular ranges of $e_{\rm r}$ (e.g.,
$0.01\lesssim e_{\rm r}\lesssim 0.1$ for BBH), but greatly reduced
for high residual eccentricities (e.g., $e_{\rm r}>0.1$ for BNSs).
In general, the BBH progenitors produce 200 and 10 times higher
GW-SGRB events than the BNS and NSBH progenitors, respectively.
Therefore, binaries with low residual eccentricities (e.g.,
$0.001\lesssim e_{\rm r}\lesssim 0.1$) and high total masses will
easier to be detected by aLIGO. However, only a small fraction of
BBHs could be SGRB progenitors (if they can produce SGRBs), because
the predicted GW-SGRB event rate (60$\sim$100 per year) is too high
compared with the recent observations, unless they merge with high
residual eccentricities (e.g., $e_r>0.7$).

\end{abstract}


\section{Introduction}

Binary compact objects, such as binary neutron stars (BNSs), neutron
star-black hole binaries (NSBHs), and binary black holes (BBHs) are
expected to be the best candidates for the sources of gravitational
waves \citep[GWs; e.g., ][]{Abbott16c, Abbott16d}, which are
expected to be detected by LIGO \citep[e.g.,][]{Abbott08, Abbott09,
Aasi15} and Virgo \citep[e.g.,][]{Acernese08,  Acernese15}.
Theoretically, short GRBs (SGRBs), with duration time $T_{90}< $ 2 s
\citep{Kouveliotou93}, are believed to originate from the mergers of
BNSs or NSBH binaries \citep{Paczynski86, Eichler89, Paczynski91,
Meszaros92, Narayan92}. Observations of SGRBs, such as the
non-detection of supernova associations, large offsets of their
locations in the host galaxies, and a possible kilonova association
\citep[e.g., GRB 130603B;][]{Tanvir13}, support the hypothesis of
the coalescing model \citep{Berger14}. Except for the
electromagnetic radiation (EMR), the coalescences of compact
binaries could generate strong GWs in the sensitive frequency band
of the ground-based GW detectors \citep{Thorne87}. On August
17 2017, the Advanced LIGO (aLIGO) and Advanced Virgo (aVirgo)
discovered the GW170817 from a binary neutron star inspiral
\citep{Abbott17b}, and the associated GRB 170817A was observed 1.7 s
after the coalescence time \citep{von Kienlin17, Connaughton17,
Goldstein17a, Goldstein17b, Savchenko17b, Savchenko17c, Abbott17c}.
This observation directly proved the coalescing model for SGRBs. It
is quite expected that more observations of GWs associated with
SGRBs (GW-SGRB) will uncover the nature of SGRB central engine
\citep{Aasi14, Regimbau15}.

Except for BNSs and NSBHs, recent observations show that
BBHs may also be SGRB progenitors
\citep[e.g.,][]{Connaughton17,Verrecchia17}. On September 14, 2015,
aLIGO detected the first transient GW event called GW150914
\citep{Abbott16a}, which was produced by the final in-spiral and
ring-down phases of a BBH system with component masses of
$m_1=36.2^{+5.2}_{-3.8}\msun$ and $m_2=29.1^{+3.7}_{-4.4}\msun$
\citep{Abbott16c}. The \emph{Fermi} gamma-ray burst monitor (GBM)
observed a weak transient source above 50 keV, 0.4~s after the GW
event, which could be a possible weak SGRB \citep{Connaughton16,
Savchenko16}. The second GW event, GW151226, was observed by the
twin detectors of aLIGO on December 26, 2015 \citep{Abbott16b}. The
inferred initial BH masses are $14.2_{-3.7}^{+8.3}\msun$ and
$7.5_{-2.3}^{+2.3}\msun$, and the final BH mass is
$20.8_{-1.7}^{+6.1}\msun$. However, no EMR was observed by the
following observations \citep{Adriani16, Cowperthwaite16}. The third
GW event, GW170104, was observed by the twin advanced detectors of
LIGO on January 4, 2017 \citep{LIGO17a}. The inferred black hole
masses are $31.2^{+8.4}_{-6.0}\msun$ and $19.4^{+5.3}_{-5.9}\msun$,
and the final black hole mass is $48.7^{+5.7}_{-4.6}\msun$. The
Fermi Gamma-ray Burst Monitor (GBM) and Large Area Telescope (LAT)
observations found no electromagnetic counterparts for GW170104
\citep{Fermi17}. Interestingly, the data from the Mini-Calorimeter
(MCAL) on board AGILE satellite shows a weak EMR event before
GW170104 (E2), and the significance for a temporal coincidence is
3.4 $\sigma$ \citep{Verrecchia17}. However, observation from
the INTErnational Gamma-Ray Astrophysics Laboratory (INTEGRAL)
challenges their result \citep{Savchenko17a}. The fourth GW
event of GW170814 was observed by a three-detector network, and the
inferred masses of BHs are $30.5^{+5.7}_{-3.0}\msun$ and
$25.3^{+2.8}_{-4.2}\msun$ \citep{Abbott17d}. Still, no obvious
optical counterparts were detected \citep[e.g.,][]{Arcavi17}. For
the fifth GW event of GW170608, the inferred BH masses are
$12^{+7}_{-2}\msun$ and $7^{+2}_{-2}\msun$ \citep{Abbott17e}, and no
EMR counterparts were reported. If BBHs are confirmed to be the
progenitors of SGRBs, then SGRB model will need to be modified
\citep[e.g., ][]{Loeb16, Zhang16}, because generally mergers of BBHs
can not produce GRBs.

The coalescence processes of compact binaries could be divided into
three phases: in-spiral, merger and ring-down
\citep[e.g.,][]{Flanagan98, Bartos13}. For the in-spiral phase, the
emitted GW frequencies are in the most sensitive band of LIGO/Virgo
and are of the most interest, which have been investigated by
numerous authors with circular orbit \citep[e.g.,][]{Tumlinson00,
Regimbau08, Rosado11, Clark15}. BNSs in our Milky Way Galaxy, whose
progenitors are the massive binary stars \footnote{Usually,
the component masses of the massive binaries are assumed within
$8\sim40\msun$ \citep[e.g.,][]{Bhattacharya91, Portegies98}},
always have low-kick velocities \citep[$\lesssim 50 \km\s^{-1}$ ;
][]{Bhattacharya91, Portegies98, Dewi05}, and their orbits are
nearly circular before the GWs enter the detection band of
LIGO/Virgo \citep[e.g., binary pulsar PSR 1913 + 16 is expected to
enter the detection frequency of 15 Hz with the residual
eccentricity of $10^{-6}$; ][]{Peters63,Kalogera01,Oslowski11}.
However, some SGRBs detected in or near the elliptical
galaxies always have large spatial offsets, which cannot be
attributed by the kick velocity. For example, GRB 050509b with an
offset of $40\pm13 \kpc$ implies a kick velocity of $200\km
\s^{-1}\lesssim v \lesssim 600\km \s^{-1}$
\citep[e.g.,][]{Grindlay06}. This is inconsistent with the
Hulse-Taylor binary pulsar \citep{Hulse75} or the BNSs known in our
Galaxy that were formed by the evolution of massive binary systems
\citep[][]{Bhattacharya91, Portegies98}. These are all probably
low-kick velocity ($\lesssim50\km\s^{-1}$) systems and thus are
expected to remain within the central potential of their parent
galaxies \citep{Dewi05}. Additionally, the birth kicks of black
holes in low-mass X-ray binaries (LMXB) also do not require to
exceed $100~{\rm km/s}$ \citep{Mandel16}. Numerical simulations show
that this kind of SGRBs could be originated from the mergers of BNSs
formed by dynamical captures in core-collapsed globular clusters
\citep{Grindlay06}. In these globular clusters, $n$-body
interactions may result in the mergers of binaries with sizable
eccentricities \citep[e.g., ][]{Anderson90}. Nearly 10\%$\sim$ 30\%
of SGRBs may formed by this mechanism \citep{Grindlay06}. One
possible evidence for binary formed by dynamical processes is the
millisecond pulsar M15-C, whose companion is a neutron star
\citep{Anderson90}. The formation mechanism could be a NS exchange
interaction with a cluster LMXB, leading to the production of a
NS-NS binary with an orbital eccentricity of $e=0.68$. Observations
of millisecond pulsar PSR J1903+0327 with a main-sequence star
companion in a high eccentric orbit ($e=0.44$) may also support the
dynamical capture mechanism \citep{Champion08}. Observations of
LMXBs or X-ray sources by \emph{Chandra} and \emph{XMM-Newton}
suggest that the stellar encounters are common in globular clusters
\citep{Gendre03, Heinke03, Pooley03}. The dynamically captured NSBHs
would also allow mergers with high eccentricities. The merger rate
in globular clusters peaks at $8-25\yr^{-1}\Gpc^{-3}$ for fiducial
systems \citep{Lee10} and $30-100\yr^{-1}\Gpc^{-3}$ for linearly
extrapolation \citep{Stephens11}. Therefore, mergers of eccentric
NSBHs could also contribute significantly to SGRB population.
\cite{OLeary09} found that the the merger rate of the eccentric BBHs
in galactic nuclei detectable by LIGO is $\sim1-10^2 \yr^{-1}$, and
the actual merger rate is likely $\sim 10$ times higher. If
BBHs could be SGRB progenitors, a significant fraction of SGRBs
could be contributed by the mergers of eccentric BBHs. Numerical
simulations show that these dynamically captured systems will merge
with a sizable eccentricity and enter the observation frequency
window of LIGO/Virgo \citep{East12, East13, Gold12, Samsing14}.
Another mechanism that could lead to high-eccentric-binary merger
within Hubble time is the the Kozai oscillation in the triple system
in galactic nuclei \citep{Thompson11, Antonini12}, which could
contribute nearly 10\% of the coalesce rate to SGRBs
\citep{Abadie10}.

Given that a large fraction of SGRB progenitors could be the
eccentric compact binaries, and the effects of the eccentricity on
GW signal search are non-negligible \citep[e.g., $e\gtrsim
0.02$;][]{Brown10, Huerta13, Coughlin15}, we will study the effects
of the eccentricity on both the SGWB and GW-SGRB co-detection rate.
In fact, SGWBs from the eccentric orbit binaries have been
studied by many previous works, such as SGWBs from the eccentric
BNSs in our Galaxy \citep{Ignatiev01}, SGWBs from the eccentric
supermassive BBHs \citep{Enoki07}, SGWBs from the eccentric
Population III binaries \citep{Kowalska12}, and SGWBs from the
long-lived eccentric BNSs \citep{Evangelista15}. In this paper,
three types of SGRB progenitors (including BNSs, NSBHs, and BBHs)
are considered as the GW sources. SGWBs are generated by these
progenitors including triggered and untriggered SGRBs. For the
co-detection rate of GW-SGRBs, only GW events with signal to noise
ratio (SNR) higher than 8 that triggered the GRB detector are
considered. While calculating the SNR, we only considered the
in-spiral phase for low mass systems (like BNS or NSBH), because GW
frequencies from the merger and ring-down phases are higher than the
sensitive frequency band of aLIGO/aVirgo. But for high mass systems
(like BBHs), three phases are all considered. Only the gamma-ray
radiation happen to point at earth could be observed as GRBs.
Therefore, just a small fraction of GW events associate with SGRBs
could be observed, because the opening angles of SGRBs are always
small \citep{Fong14}. However, there are many advantages for GW-SGRB
observations, e.g., improving the detection probability of GW signal
in comparison with an arbitrary stretch of data \citep{Williamson14,
Bartos15, Clark15}. Therefore, GWs associate with SGRBs will be an
optimal direction for the detection of GW signals. In our
calculation, all SGRBs are assumed to originate from the mergers of
eccentric compact binaries. It is perhaps even more likely that not
all SGRBs are generated from the eccentric compact binaries. We thus
expect that our predictions for the event rates are optimistic.

\section{SGRB rate}
\begin{figure}
\centering\resizebox{0.45\textwidth}{!}{\includegraphics{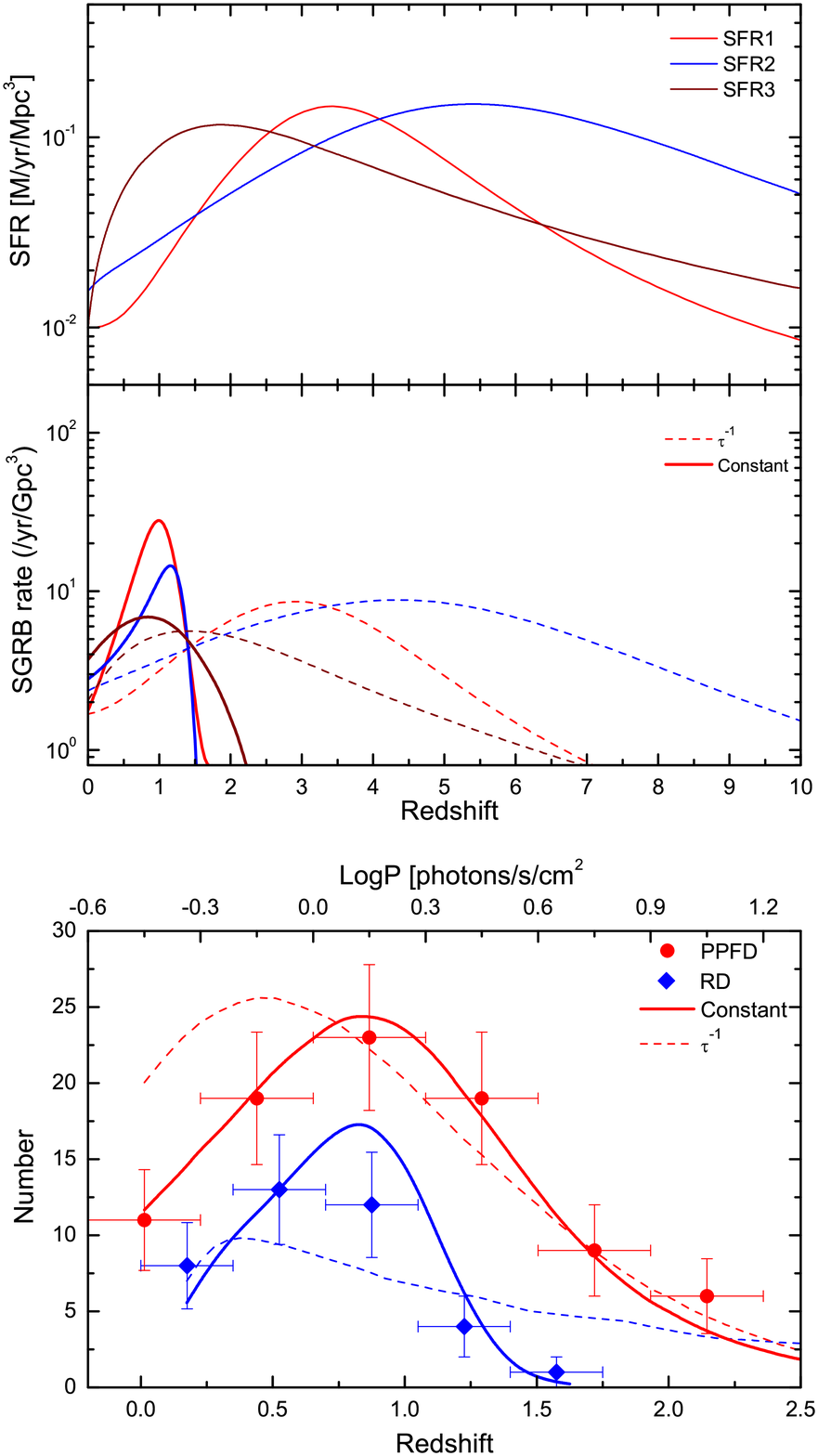}}
\caption{Top: Cosmic star formation rates used in this paper. SFR
from \protect\cite{Robertson12}  and Wang (2013) (SFR1; red line),
SFR from \protect\cite{Springel03} (SFR2; blue line), and SFR from
\protect\cite{Fardal07} (SFR3; wine line). Middle: The corresponding
SGRB rates with a power law time delay distribution ($P_\tau\propto
\tau^{-1}$; dashed lines) and a constant time delay distribution
(solid lines), respectively. Bottom: The best-fitting results of the
peak photon flux distribution (PPFD; top label) and the redshift
distribution (RD; bottom label) for SFR1.}\label{figure 1}
\end{figure}

In this section, we will constrain the SGRB burst rate. Following
\cite{Tan15} (where long GRBs are considered), we will fit the
observed SGRB peak photon flux distribution (PPFD; in unit of $\rm
photons/\s/\cm^{2}$) and redshift distribution (RD) simultaneously
to derive the luminosity function (LF) and the SGRB rate. The peak
photon flux (PPF) and redshift data are taken from the \emph{Swift}
archive
\footnote{http://swift.gsfc.nasa.gov/docs/swift/archive/grb\_table.}.
We also include some probable SGRBs mentioned in literature
\citep{Dietz11,Kopac12,Berger14}. Finally, we obtained 87 SGRBs with
PPFs (in the energy band of 15-150 \kev) and 38 SGRBs with
redshifts. The PPFD and RD are shown in the bottom panel of Figure
\ref{figure 1} by red circles and blue diamonds, respectively. The
error bars along $y$-axis are the statistical errors (i.e., square
root of the number in each bin $\Delta N=\sqrt{N}$), which
correspond to 68\% Poisson confidence intervals for the binned
events. The error bars of $x$-axis represent the bin size.

In order to derive the SGRB burst rate, we firstly introduce the
parameters as follows. $\Phi_{P}(L)$ is the LF with
$L_{\min}=10^{49}\erg\s^{-1}$ and $L_{\max}=10^{55}\erg\s^{-1}$ for
normalization. $\eta(P_E)$ is the flux triggering efficiency of
\emph{Swift} \citep{Lien14},
\begin{eqnarray}
\eta(P_E)={a(b+c P_E/P_{E,0})\over (1+P_E/d P_{E,0})} \label{trigger
function}
\end{eqnarray}
for $P_E>5.5\times 10^{-9}\erg\s^{-1}\cm^{-2}$, and below this range
the function equals to zero (which is suggested to be better than a
single detection threshold \citep{Howell14, Tan15}). The parameters
are as follows: a=0.47, b=-0.05, c=1.46, d=1.45 and
$P_{E,0}=1.6\times 10^{-7}\erg\s^{-1}\cm^{-2}$ (Howell et al. 2014).

 $P_{\rm E}$ is the
peak energy flux in units of $\erg/\s/\cm^2$, which could be related
to the peak photon flux ($P$) by
\begin{eqnarray}
P_{\rm E}(P,z)={P \over (1+z)}{\int_{15(1+z)}^{150(1+z)}E S(E){\rm
d}E\over \int_{15(1+z)}^{150(1+z)}S(E){\rm d}E}\kev,
\end{eqnarray}
where $S(E)$ is the prompt spectrum of SGRB, which is well modeled
by the band function \citep{Band93}. The spectral indices are $-0.5$
and $-2.3$ below and above the peak energy, respectively. The peak
energy could be derived by the $E_{\rm p}-L_{\rm p}$ relation
proposed by \cite{Tsutsui13},
\begin{eqnarray}
L_{\rm p}=10^{52.29\pm0.066}\erg\s^{-1}\left[{E_{\rm p}(1+z)\over
774.5\kev}\right]^{1.59\pm0.11},
\end{eqnarray}
with the linear correlation coefficient is 0.98 and the chance
probability is 1.5$\times 10^{-5}$ \citep{Tsutsui13}. Zhang \& Wang (2017)
found a similar correlation using more SGRBs \citep{Zhang17}.

The peak photon flux in the detector can be transformed into the
peak luminosity in a straightforward way if the redshift is known.
We describe it as
\begin{eqnarray}
L_{\rm p}=4\pi d_{\rm L}(z)^2 P_{\rm E}(P) \mathcal{K}(z).
\end{eqnarray}
Here $\mathcal{K}(z)$ is the correction factor, which is to convert
the observed energy band of $15-150 \kev$ to the rest frame band of
$1-10^4 \kev$. We describe it by
\begin{eqnarray}
\mathcal{K}(z)={\int_{15(1+z)}^{150(1+z)}E
S(E)dE\over\int_{1}^{10^4}E S(E)dE}.
\end{eqnarray}

For the luminosity function, a broken power-law form is assumed,
which is given by
\begin{eqnarray}
\Phi_{\rm p}(L)\propto\left\{~\begin{array}{ll}\left({L\over
L_b}\right)^{-\alpha},~~~~& L\leq L_b,\,\\
\left({L\over L_b}\right)^{-\beta},~~~~&L>L_{b},\,\end{array}\right.
\label{LF},
\end{eqnarray}
where $L$ is the peak luminosity in the energy band of $1-10^4\kev$,
and $\alpha$ and $\beta$ are the power law indices below and above
the break luminosity $L_{\rm b}$. If $\alpha$ equals to $\beta$, the
LF transforms into a single power law form.

$R_{\rm SGRB}(z)$ is the observed SGRB rate, which could be related
to the star formation rate (SFR) by
\begin{eqnarray}
R_{\rm SGRB}(t)\propto \int R_{\rm SFR}(t-\tau)P_{\tau}({\tau}){\rm
d} \tau,
\end{eqnarray}
where $R_{\rm SFR}(t)$ is the SFR and $t$ is the cosmic time
corresponding to redshift $z$. $\tau$ is the time delay between the
star formation and the occurrence of SGRB, and $P_{\tau}({\tau})$ is
the probability distribution function of $\tau$.

For the parameters defined above, the expected number of SGRBs with
observed PPF between $P_1$ and $P_2$ that triggered BAT onboard
\emph{Swift} can be expressed by

\begin{eqnarray}
N(P_1, P_2)={\Delta\Omega\over 4\pi} T
\int^{\infty}_{0}\int^{L_{\max}}_{L_{\min}}\eta(P_{\rm E})\Phi_{\rm
p}(L) R_{\rm SGRB}(z){\rm d}L{{\rm d}V(z)\over 1+z}, \label{expected
P}
\end{eqnarray}
with $P_{\rm E}=P_{\rm E}( P,z)$, $L=L(z, P_{\rm E})$. The expected
number of SGRBs within redshift range of $z_1<z<z_2$ is given by
\begin{eqnarray}
N(z_1, z_2)={\Delta\Omega\over 4\pi} T
\int^{z_2}_{z_1}\int^{L_{\max}}_{L_{\rm min}}\eta(P_{\rm
E})\Phi_{\rm p}(L)R_{\rm SGRB}(z){\rm d}L{{\rm d}V(z)\over 1+z},
\label{expected z}
\end{eqnarray}
where $(\Delta\Omega/4\pi)\sim0.1$ is the field view of BAT,
$T\sim10 ~\rm yrs$ is the observation time period, $\eta(P_{\rm E})$
is the triggering function, ${\rm d}V(z)$ is the comoving volume
element, $1/(1+z)$ accounts for the time dilation.

There is an agreement that SGRB rates do not trace the SFRs directly
but with a time delay. Usually, a power-law time delay distribution
($P_{\tau}\propto \tau^{-1}$) and a lognormal time delay
distribution are considered. However, it is found that a power-law
time delay distribution is disfavored but a lognormal time delay
distribution with $\tau \gtrsim3~ {\rm Gyr}$ is consistent with the
observations \citep{Ando04, Guetta06, Guetta09, Hao13, Wanderman15}.
For the lognormal distribution, \cite{Wanderman15} found that the
width of the distribution is $\sigma\lesssim0.2$ at $68\%$
confidence level, which corresponds to a very small spread by a
factor of $\lesssim 1.2$ in the time delays. Therefore, a constant
time delay distribution is assumed in our work for simple, which can
be expressed by $R_{\rm SGRB}(t)\propto R_{\rm SFR}(t-\tau)$. Here,
we consider three SFR models: SFR derived from long GRBs
\citep[SFR1; red line; ][]{Robertson12,Wang13}, SFR derived from
cosmological smoothed particle hydrodynamics numerical simulations
\citep[SFR2; blue line; ][]{Springel03}, and SFR derived from
observations \cite[SFR3; wine line; ][]{Fardal07}, which are shown
in the top panel of figure \ref{figure 1}.

The free parameters of $\alpha$, $\beta$, $L_{\rm b}$, and $\tau$
are fitted jointly with PPFD and RD. Firstly, we give an arbitrary
set of values for the free parameters. Then, for each set of the
parameters, we calculate the $\chi^2$ values (Pearson's chi-squared
value) for both PPFD and RD. The total $\chi^2$ is assumed to be the
linear combination of $\chi^2_{\rm PPFD}$ and $\chi^2_{\rm RD}$. The
best-fitting parameters are derived by minimizing the global
$\chi^2$. Finally, we show the best-fitting parameters with
$1\sigma$ errors and $Q$ values (the probability to find a new total
$\chi^2$ exceeding the current one) for different SFR models in
Table \ref{table 1}. The corresponding SGRB rates are shown in the
middle panel of figure \ref{figure 1}. For comparison, we fitted the
PPFD and RD for SFR1 with a constant time delay distribution and a
power law time delay distribution ($P_{\tau}\propto \tau^{-1}$),
simultaneously. The best fitting results are shown in the bottom
panel of figure \ref{figure 1}. It is obvious that the constant time
delay distribution (thick lines) fits the observations quite well,
while the power-law time delay distribution (thin lines) fits quite
bad, because it produces too many dim SGRBs at high redshifts. The
same situation occurs in the other two SFR models.
%SGRBs.

%The fitting results are shown in the bottom panel of Figure
%\ref{figure 1}.

%The error bars for the parameters are obtained by setting
%$\chi^2=\chi^2_{\min}+3.5$, which corresponds to the parameters
%within 68.3\% confidence level.

% \textbf{For comparison, we
%also calculated the SGRB rates for different SFR models. As shown in
%the upper panel of Figure \ref{figure 1}, the SGRB rates are nearly
%the same for different SFRs (with different constant time delay),
%because the SGRB rate is only determinate by the PPFD and RD. It is
%obviously that a power-law time delay distribution could not
%reproduce the observations, because it predicts more high redshift
%SGRBs.}

\begin{table}
\centering \caption{The best-fitting results for different SFR
models with the constant time delay distribution of $P_{\tau}$.}
\begin{tabular}{cccccc}
\hline\hline
Model  & $R_{\rm SGRB}(0) (\yr^{-1}\Gpc^{-3}$) & $\tau(\Gyr)$ &$\alpha=\beta$&$\chi^2$& $Q$ \\
\hline  SFR1&$1.77^{+0.4}_{-0.31}$&$3.65^{+0.28}_{-0.39}$&$1.6^{+0.06}_{-0.06}$&$6.18$&0.52\\
SFR2&$2.8^{+0.62}_{-0.46}$&$3.88^{+0.18}_{-0.45}$&$1.6^{+0.05}_{-0.05}$&5.89&0.55\\
SFR3&$3.7^{+0.85}_{-0.61}$&$2.52^{+0.98}_{-1.3}$&$1.6^{+0.05}_{-0.05}$&2.38&0.94\\
\hline\hline
\end{tabular}\begin{flushleft}
Notes:The best-fitting parameters for three SFR
models. $R_{\rm SGRB}(0)$ is the observed local SGRB rate, $\tau$ is
the constant time delay between SFR and SGRB rate, $\alpha$ and
$\beta$ are the power law indices of the LF in equation (\ref{LF}),
$\chi^2$ is the global chi-squared value, and Q represents the
probability to find a new total $\chi^2$ exceeding the current one.
\end{flushleft}\label{table 1}
\end{table}

Given that SGRBs are originated from the mergers of compact
binaries, we could be able to predict the total merger rate of their
progenitors, where the beaming effect should be considered.
Observations of jet breaks in SGRB afterglows can be used to
constrain the jet opening angles \citep{Sari99}. Some SGRBs with jet
break observed have been used to derive the opening angles, e.g.,
GRB 051221A ($7^\circ$), GRB090426 ($5^\circ-7^\circ$), GRB111020A
($3^\circ-8^\circ$), GRB 130603B ($4^\circ-8^\circ$) (please refer
to \cite{Fong14} for review). However, there are many SGRBs without
jet break observation in their afterglows, and only the lower limits
of the opening angles could be given, e.g., GRB 050709 with
$\theta_{j}>15^\circ$. Therefore, we use $\theta_{j}=5^\circ$ and
$\theta_{j}=20^\circ$ as the lower and upper limits for the opening
angle to infer the total binary merger rate
\begin{eqnarray}\label{merger rate}
R_{\rm merger}\propto R_{\rm SGRB}/(1-{\rm{cos}}~\theta_{j}),
\end{eqnarray}
where we assume that all SGRBs are produced by the mergers of
compact binaries.

% GRB 050724A, GRB 101219A, GRB 111117A, GRB
%120804A, GRB 050709 and GRB 081226A
\section{SGWB from eccentric orbit binaries}

SGWBs from the circular orbit binaries have been well studied
\citep{Tumlinson00, Regimbau08, Rosado11, Wu12, Zhu13, Clark15}.
Considering the progenitors of SGRBs could be eccentric compact
binaries, we will study the effects of eccentricity on SGWB in this
section \citep{Ignatiev01,Enoki07,Kowalska12,Evangelista15}, and
compare it with the ground-based detectors \citep{Kowalska12}.

Different from the circular orbit binaries, the velocity on the
eccentric orbit changes over its period, and the instantaneous
orbital frequency also varies. Therefore, binaries will radiate GWs
across some range of frequencies but not at one particular
frequency. For the in-spiral phase, the instantaneous spectrum of
GWs emitted by the eccentric orbit binaries in the source rest frame
can be describe by \citep[e.g., ][]{Peters63, Kowalska12}
\begin{eqnarray}
\label{dEdf} {{\rm d}E_{\rm i}\over {\rm d} f^{n}_{\rm gw}}={\pi
\over 3}{1\over G}\left({4\over n^2}\right)^{1/3} {(G M_{\rm
chirp})^{5/3}\over (f^n_{\rm gw}\pi)^{1/3}}{g(n,e)\over \Psi(e)},
\end{eqnarray}
and the spectrum at a special frequency could be described by
\begin{eqnarray}
{{\rm d}E_i \over {\rm d}f_{\rm gw }}=\sum_{n=2}^{\infty}
\delta(f_{gw}-{f_{\rm gw}^{n}}){{\rm d}E_i\over {\rm d} f^{n}_{\rm
gw}}\bigg|_{f_{gw}=f(1+z)},\label{inspiral}\end{eqnarray}

\noindent where $M_{\rm chirp}=\mu^{3/5}M^{2/5}$ is the chirp mass
of the binary, the total mass $M$ and the reduced mass $\mu$.
$f^{n}_{\rm gw}=n f_{\rm orb}$ is the eccentric orbit binary emits
GWs of harmonics of the orbital frequency with $n\geq2$, and $f$ is
the GW frequency in the observer's frame. While $e=0$, the above
equation reduces to the circular orbit case, since $\Psi(e=0)=1,
g(n=2, e=0)=1$, and $g(n\ne2, e=0)=0$. Here $g(n,e$) and $\Psi(e)$
are described as follows
\begin{eqnarray}
\label{gne} g(n,e) & = & \frac{n^4}{32} \left\{ [
J_{n-2}(ne)-2eJ_{n-1}(ne)+\frac{2}{n}J_n(ne)
\right.  \nonumber\\
& &   +2eJ_{n+1}(ne)-J_{n+2}(ne) ]^2 \nonumber\\
& & +(1-e^2) \left[J_{n-2}(ne)-2J_n(ne)+J_{n+2}(ne) \right]^2 \nonumber\\
& &\left.  +\frac{4}{3n^2} \left[ J_n(ne) \right]^2 \right\},\\
\Psi(e) & = & \frac{1+73/24 e^2 + 37/96 e^4}{(1-e^2)^{7/2}},
\label{psi}
\end{eqnarray}
where $J_n$ are the Bessel functions.

For BBHs, GW radiated in the merger and ring-down phases are also
quite important while calculating the SGWB and SNR
\citep[e.g.,][]{LIGO17a}. For the merger phase, we assume that the
GW energy is confined to the frequency regime of $f>f_i$, where
$f_i$ ccould be taken as \citep[e.g.,][]{Kidder93, Lai96}
\begin{eqnarray}
f_{i}\sim {c^3\over6\sqrt{6}G \pi M}.
\end{eqnarray}
The merger phase will end when the waveform can be described by the
$l=m=2$ quasi-normal mode signal of a Kerr BH. The quasi-normal
ringing frequency gives the upper bound of the GW frequency radiated
in the merger phase \citep{Flanagan98}
\begin{eqnarray}
f_{q}\sim {F(a)c^3\over 2 \pi G M},
\end{eqnarray}
where $F(a)=1-0.63(1-a)^{3/10}$ and $a=0.7$ is ther dimensionless
spin parameter of the BH \citep[e.g., ][]{Abbott16a, Abbott16b,
LIGO17a}. The total energy radiated in the merger phase is
\citep{Kobayashi03}
\begin{eqnarray}
E_{m}=\epsilon_m \left({4\mu \over M}\right)^2Mc^2,
\end{eqnarray}
where $\epsilon_m=0.05$ is the parametrization of the total energy
radiated in the coalescence. Although the energy spectrum of the
merger phase may have some features related to the dynamical
instabilities \citep{Xing94, Dimmelmeier02}, we assume a simple flat
spectrum
\begin{eqnarray}
{{\rm d}E_m\over {\rm d} f_{gw}}={E_m\over
f_q-f_i}\bigg|_{f_{gw}=f(1+z)}.\label{merger}
\end{eqnarray}

For the ring-down phase, the spectrum peaked at $f_q$ with a width
of $\triangle f\sim \tau^{-1}=\pi f_q/Q(a)$ with
$Q(a)=2(1-a)^{-9/20}$ \citep{Echeverria89}:
\begin{eqnarray}
&&\hspace{-0.6cm}{{\rm d}E_r\over {\rm d} f_{gw}}\sim
{E_rf^2\over4\pi^4f_q^2\tau^3}\nonumber\\
&&\hspace{-0.6cm}\times\left\{{1\over[(f_{gw}-f_q)^2+(2\pi\tau)^{-2}]^2}
+{1\over[(f_{gw}+f_q)^2+(2\pi\tau)^{-2}]^2}\right\}\bigg|_{f_{gw}=f(1+z)}\label{ring-down},
\end{eqnarray}
where $E_r=\epsilon_r(4\mu/M)^2Mc^2$ is the total energy radiated in
the ring-down phase. We assume $\epsilon_r=0.01$ as a nominal
parameter \citep{Kobayashi03}.

\begin{figure}
\centering\resizebox{0.45\textwidth}{!}{\includegraphics{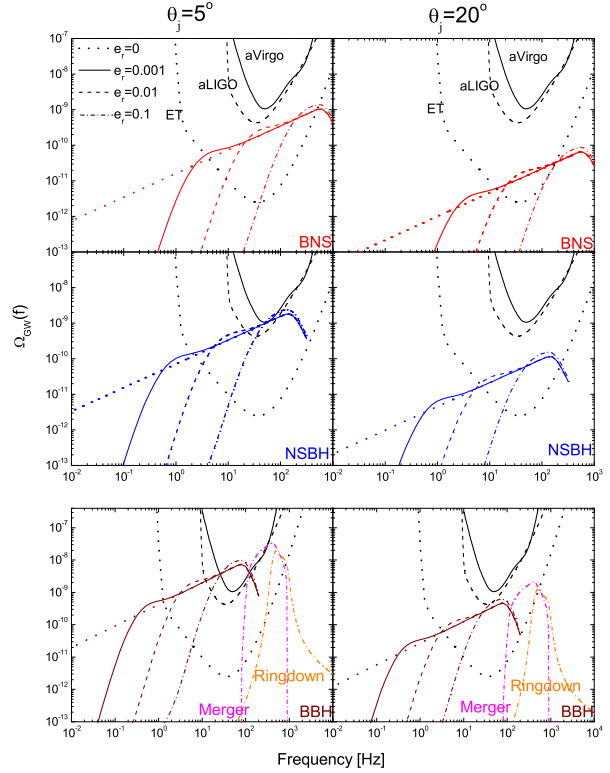}}
\caption{SGWBs from the eccentric BNSs (top panel), NSBH
binaries(middle panel), and BBHs (bottom panel). Two jet opening
angles of $\theta_j=5^\circ$ (left) and $\theta_j=20^\circ$ (right)
are considered. The residual eccentricities considered here are:
$e_r=0$ (circular case, dotted lines), $e_r=0.001$ (solid lines),
0.01 (dashed lines), and 0.1 (dash-dotted lines). For BBHs, SGWBs
from the merger (pink dash-dotted line) and ring-down phases (orange
dash-dotted line) are also calculated. The detection thresholds of
the advanced Virgo (aVirgo), aLIGO , and ET telescopes are shown as
black lines for one year of observation.}\label{figure 2}
\end{figure}

SGWB is always described by the dimensionless energy density
parameter of $\Omega_{\rm GW}(f)$, which is the present GW energy
density per logarithmic frequency interval divided by the critical
energy density of the present universe ($\rho_{\rm c}c^2$)
\citep{Phinney01},
\begin{eqnarray}
\Omega_{\rm GW}(f)={1\over \rho_c c^2}{{\rm d}\rho_{\rm gw}\over{\rm
d~ln}f},
\end{eqnarray}
where $\rho_{\rm gw}$ is the GW energy density, $f$ is the frequency
in the observer's frame, and $\rho_c={3H_0^2/ 8 \pi G}$ is the
critical energy density of the universe. For the astrophysical
origin of the GW background, $\Omega_{\rm GW}(f)$ can be given by
\begin{eqnarray}
\Omega_{\rm GW}(f)={f\over \rho_c c^3} F(f),
\end{eqnarray}
where $F(f)$ is the integrated GW flux at the observed frequency
$f$, which can be described by
\begin{eqnarray}
F(f)=\int F_s(f,z){R_{\rm merger}(z)\over 1+z}{{\rm d}V(z)},
\end{eqnarray}
$F_s(f,z)$ is the observed GW fluence of a single source
\begin{eqnarray}\label{Fs}
F_s(f,z)={(1+z)^2\over 4\pi d_L^2}{{\rm d}E\over{\rm d} f^{}_{\rm gw
}}\bigg|_{f_{gw}=f(1+z)},
\end{eqnarray}
we use equations (\ref{inspiral}), (\ref{merger}) and
(\ref{ring-down}) to calculate the SGWBs from the in-spiral, merger
and ring-down phases, respectively. Here $d_L$ is the luminosity
distance. Combining the binary merger rate derived from SGRBs of
equation (\ref{merger rate}) with equations (\ref{dEdf})-(\ref{Fs}),
we simplify $\Omega_{\rm GW}(f)$ by
\begin{eqnarray}\label{Omega}
\Omega_{\rm GW}(f)=\int^{z_{\max}}_{z_{\min}}{8\pi G \over 3
\rm{H}_0^3 c^2} {R_{\rm merger}(z)\over (1+z)}{f~\over
\varepsilon(z)} \left({{\rm d}E\over{\rm d} f^{}_{\rm
gw}}\right)\bigg|_{f_{gw}=f(1+z)} {\rm d}z,
\end{eqnarray}
here $\varepsilon(z)=\sqrt{\Omega_{\rm M}(1+z)^3+\Omega_{\Lambda}}$.
For the in-spiral phase, we set $z_{\min}=0$ and $z_{\max}=6$ for
$f<f_{\max}/(1+z_{\max})$, otherwise $z_{\max}=f_{\max}/f-1$
\citep{Rosado11, Wu12}. The maximum frequency
$f_{\max}=(1-e_r^2)^{3/4}c^3/6\sqrt{6}G \pi M$ corresponds to the
last stable orbit of the binary, with the semi-minor axis three
times the Schwarzschild radius of each star. For the merger and
ring-down phases, we set $z_{\min}=0$ and $z_{\max}=6$.

The eccentricity evolves as the orbital decay because of the GW
radiation. The evolution of the semi-major axis and eccentricity can
be described in the quadruple approximation by the differential
equations in \cite{Peters63}. The eccentricity and semi-major axis
evolve according to the two differential equations of
\begin{eqnarray}
{{\rm d}a\over{\rm d}t}&=&-{\beta\over
a^3}\Psi(e)~~~~~~~~~~~~~~~~~~~~~~~~\beta={64\over 5} {G M_1M_2 M\over c^5},\\
{{\rm d}e\over{\rm d}t}&=&-{19\over12}{\beta\over
a^4}\Theta(e)~~~~~~~~~~~~~~\Theta(e)={(1+121/304e^2)e\over(1-e^2)^{5/2}}.
\end{eqnarray}
Especially, we assume that all binaries do not reach the circular
orbit at the last stable orbits ($a_{\min}=6 G M/c^2\sqrt{1-e_r^2}$)
but with a same residual eccentricity ($e_r$). For comparison, we
give a set values of $e_r=0.001$, 0.01, and 0.1. Combing
equations (25)-(26) with the final condition of the in-spiral phase
($e_r=0.001, 0.01$, and $0.1$ at $a_{\min}$), we derive the
evolution of the eccentricity ($e$) with the semi-major axis ($a$).
Here, we do not care about the initial separation of the binary or
the initial eccentricity but stopped reversing at $a_0=10^{12} \cm$
(where $e_0\simeq1$), because GWs radiated at $a>a_0$ nearly have no
contribution to SGWB in and near the detection bands of aLIGO/aVirgo
(e.g., 0.1Hz$<f<$1000Hz).

We show the SGWBs generated by the eccentric BNSs, NSBHs, and BBHs
with two opening angels of $\theta_j=5^\circ$ (left panel; the upper
limit of SGWBs) and $\theta_j=20^\circ$ (right panel, the lower
limit of SGWBs) in Figure \ref{figure 2}. The masses of the NS and
BH are assumed to be $1.4 M_\odot$ and $10 M_\odot$
\footnote{Theoretically, the masses of the stellar-mass BHs
could range from a few solar masses to sever hundred solar masses
\citep{Heger02}. However, BH masses measured in X-ray binaries are
concentrated in $10 \msun$ \citep[e.g.,][]{Casares14}, and are
mostly $30 \msun$ in the observed BBH merger events
\citep[][]{Abbott16c, LIGO17a, Abbott17d}.}, respectively. All the
binaries are assumed to have the same residual eccentricities before
merger, e.g., $e_r=0.001$, 0.01, and 0.1. For different SGRB
progenitors, GW spectra at low frequencies are greatly suppressed.
Especially, binaries with high residual eccentricities (e.g.,
$e_{\rm r}\gtrsim0.1$ for BNSs) will radiate GWs in the
high frequency bands (e.g., $f>100  {\rm Hz}$), which even could be
out of the detection bands of aLIGO/aVirgo. It means the
non-detection of the SGWB could be resulted by the high residual
eccentricities of binaries before merger. We also calculated the
SGWBs from NSBHs and BBHs (as shown in the middle and bottom panels
of figure \ref{figure 2}), both SGWBs are higher than that of BNSs,
but the maximum frequencies of the in-spiral phase are lower,
because $f_{\rm max}$ is proportional to $M^{-1}$. The effects of
the eccentricity on SGWBs are similar for all SGRB progenitors,
e.g., suppressing the SGWB at low frequencies. Especially, SGWBs
from BBHs are the strongest, and the merger and ring-down phases
also contribute a lot to SGWBs at high frequencies. It is quite
expected that SGWBs from the merger of BBHs could be detected by
aLIGO/aVirgo in the near future.

The chosen of different SFR models nearly have no effect on SGWBs,
which could be clarified if we change equation (\ref{Omega}) into
\begin{eqnarray}
\Omega_{\rm GW}(f)& = & \left<N\right>{8\pi G \over 3 \rm{H}_0^2
c^3}{(1+z)^2\over 4\pi d_{\rm L}^2} {f}\left({{\rm d}E\over{\rm d}
f^{}_{\rm
gw}}\right)\bigg|_{f_{gw}=f(1+z)}\nonumber \\
&\approx&{f\over \rho_c c^3} F_s(f,z)\left<N\right> ,
\end{eqnarray}
where $\left<N\right>$ is the total SGRB rate, which is only
determinated by the observed SGRB number. $F_s(f,z)$ is the GW
fluence from a single source, which is slightly affected by the
redshift. In fact, the redshift effect is negligible, because SGRBs
are concentrated in a narrow redshift range of $z\sim 1$.

\section{GW-SGRB co-detection rate from eccentric orbit binaries}
The co-detection of GWs associate with SGRBs could reduce the search
time and increase the detection sensitivity while comparing with the
all-sky, all-time search \citep{Bartos15, Clark15}. The
co-detected events are expected to almost face on, and the search
time of the GW events could be a few seconds around the burst of
SGRBs \citep{Aasi14}. Therefore, we use the designed sensitivity
cure of aLIGO to calculate the detection distance of the GW events
\citep{Abbott16e} and assume an all sky field of view for SGRB
observations.

%The noise curve for aLIGO is given by the LIGO document T0900288-v3
%\footnote{See the web site
%http://dcc.ligo.org/LIGO-T0900288/public}, with the target design
%sensitivity given by the ZERO\_DET\_high\_P.txt curve and
%$f_{\min}=10 {\rm Hz}$.

\begin{figure}
\centering\resizebox{0.45\textwidth}{!}{\includegraphics{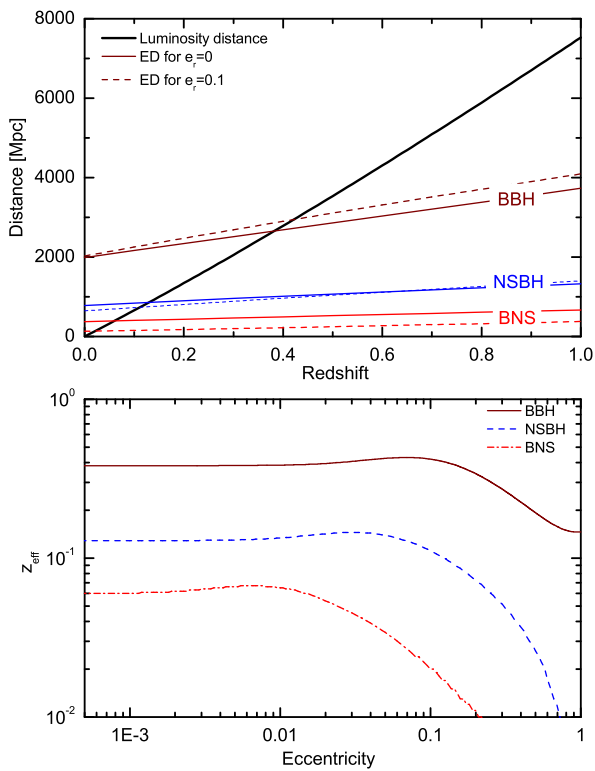}}
\caption{Top: The effective distance (ED) of the GW events (thin
lines) and the luminosity distance of SGRBs (thick line). Three SGRB
progenitors are considered. The thin solid and dashed lines are for
binaries with residual eccentricities of $e_r=0$ and $e_r=0.1$,
respectively. Bottom: The maximum redshift where SGRB could be
detected with GW signal. }\label{figure 3}
\end{figure}

To determinate whether a GW event is detectable or not, we need to
calculate its signal to noise ratio (SNR). For one single event, the
SNR could be written as \citep[e.g., ][]{Regimbau15}
\begin{equation}
\rho^2 = 4 \int_0^\infty \frac{|\tilde{h}_{+}F_{+}
+\tilde{h}_{\times} F_{\times}|^2}{S_{n}(f)}\, {\rm d}f,
\end{equation}
where $f$ is the observed GW frequency, $\tilde{h}_{+}$ and
$\tilde{h}_{\times}$ are the Fourier transforms of the GW strain
amplitudes, $F_{+}$ and $F_{\times}$ are the antenna response
functions of the detector \citep{Thorne87}, and $S_{\rm n}(f)$ is
the one-sided noise power spectral density of aLIGO \citep[e.g.,
][]{Flanagan98}. Combing equations (\ref{inspiral}), (\ref{merger}),
and (\ref{ring-down}) in this paper with equation (3) in
\cite{Regimbau15}, we could derive the SNR for the eccentric orbit
binary mergers
\begin{equation}
\rho^2 = {{5\over 2}} {G(1+z)^2{\cal F}^2 \over c^3 \pi^2 d_{\rm
L}^2}\int_{f_{\rm min}}^{f_{\rm max}\over 1+z} {f^{-2}\over
S_{n}(f)}\left({{\rm d}E\over {\rm d}f_{\rm
gw}}\right)\bigg|_{f_{gw}=f(1+z)}{\rm d}f.
\end{equation}
The factor
\begin{equation}
{\cal F}^2 = {(1+{\rm cos}^2\iota)^2\over 4}F^2_{+}+{\rm cos}^2\iota
F^2_{\times}
\end{equation}
characterizes the detector response. Here $\iota$ is the inclination
angle, and only the inclination angle within the SGRB opening angle
could be observed as GW-SGRB event. Therefore, we have ${\rm
cos}~\theta_{j} \le{\rm cos}~\iota\approx 1$.

To calculate the detectable volume of aLIGO, we need to know the
detectable distance of the source, here we adopted the so-called
effective distance (ED), which is related to $d_{\rm L}$ through
$D_{\rm eff}=d_{\rm L}/{\cal F}$\citep{Allen12}. When averaging
 $F_{+}$ and $F_{\times}$ over the uniformly distributed $\theta$ (the
right ascension of the source), $\phi$ (the declination of the
source) and $\psi$ (the polarization angle), we obtain
$\left<F_{+}^2\right>=\left<F_{\times}^2\right>={\rm sin}^2\zeta/5$,
where $\zeta=90^{\circ}$ is the opening angle between two arms of
aLIGO. Therefore, the detectable distance could be written as
\begin{eqnarray}
D_{\rm eff}={1+z\over \rho}\left({5 G\over 2 \pi^2
c^3}\right)^{1/2}\sqrt{\int_{f_{\min}}^{f_{\max}\over
1+z}{f^{-2}\over S_{n}(f)}\left({{\rm d}E\over {\rm d}f_{\rm
gw}}\right) \bigg|_{f_{gw}=f(1+z)} {\rm d} f}.\label{deff}
\end{eqnarray}
In our calculation, the threshold SNR is set to be $\rho=8$
\citep{Abbott16e}. Specially, we should note that only GWs from the
in-spiral phase are used to calculate the SNR for BNS and NS-BHs,
because GWs from the merger and ring-down phases are out of the
sensitive band of aLIGO/aVirgo. For BBHs, three phases are
considered to calculate the SNR.

Only SGRBs within the detectable distance of aLIGO could be observed
as the GW-SGRB events, therefore, ED should be greater than the
luminosity distance. As shown in the top panel of figure \ref{figure
3}, only SGRBs with redshifts smaller than the cross point of ED and
$d_{\rm L}$ could be observed as the GW-SGRB events. We found that
ED is slightly dependent on the redshift for low mass binaries
(e.g., BNSs and NSBHs), but more dependent for high mass binaries
like BBHs. To study the effect of the eccentricity on GW-SGRB
co-detection rate, we calculated the maximum redshift-eccentricity
relations for different binary systems, which are shown in the
bottom panel of figure \ref{figure 3}.

%Considering that ED also dependents on the residual eccentricity
%($e_{\rm r}$) of the binaries, we calculated the maximum ED for
%different $e_{\rm r}$, and recording the corresponding maximum
%redshift $z_{\rm eff}$. Then, we derived the relations between the
%maximum redshift $z_{\rm eff}$ and $e_{\rm r}$ for different SGRB
%progenitors, as shown in the bottom panel of figure

\begin{figure}
\centering\resizebox{0.45\textwidth}{!}{\includegraphics{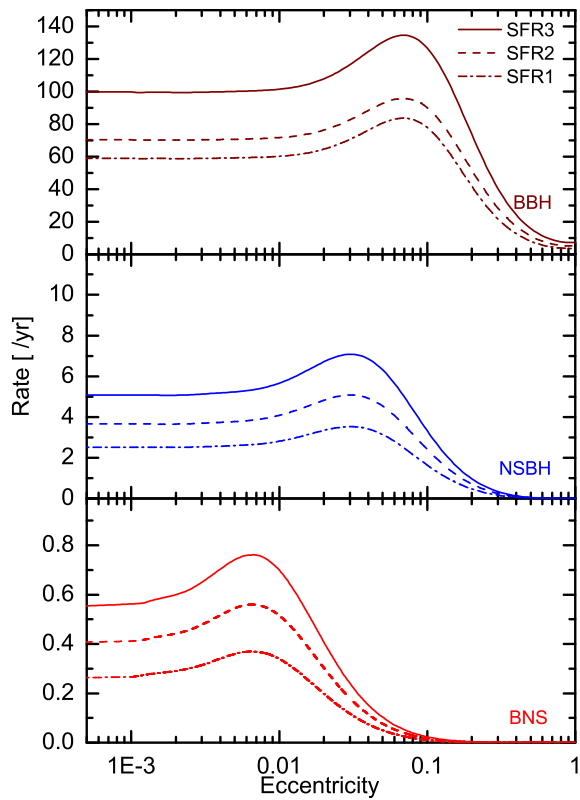}}
\caption{The co-detection rate of GW-SGRB event versus the residual
eccentricity. Three types of SGRB progenitors are considered: BBHs
(wine lines), NSBN binaries (blue lines), and BNSs (red lines). The
dash-dotted (SFR1), dashed (SFR2), and solid lines (SFR3) represent
three SFR models used in the paper.}\label{figure 4}
\end{figure}

Finally, the GW-SGRB co-detection rate is a function of the observed
SGRB rate $R_{\rm SGRB}(z)$ and the detectable volume of the GW
event, which can be described by
\begin{eqnarray}
R_{\rm co}(e_{\rm r})&=&\int_{0}^{z_{\rm eff}(e_{\rm r})} {R_{\rm
SGRB}(z)\over (1+z)} {dV(z)},\label{Rco}
\end{eqnarray}
where $R_{\rm SGRB}(z)$ is the observed SGRB rate derived in section
2, $dV(z)$ is the detectable volume of the GW event. In figure
\ref{figure 4}, we show the GW-SGRB co-detection rates as a function
of the residual eccentricity. Three cases such as BNSs (red lines),
NSBHs (blue lines) and BBHs (wine lines) are considered. We found
that the GW-SGRB co-detection rates are greatly reduced for binaries
with high residual eccentricities ($e_{\rm r}\gtrsim0.2$) while
comparing with the circular case. Therefore, much more time will be
needed for the observation of these events. For example, it will
cost 2.6 years (for SFR3) to observe one event for BNSs with
$e_r=0.02$, but 1.8 years (for SFR3) for the circular case.
Interestingly, the GW-SGRB co-detection rates could be $\sim1.4$
times higher than the circular case within some particular ranges of
$e_r$ (shown as the bumps in the figure), depending on the masses of
the binaries, e.g., $0.0013<e_{\rm r}<0.014$ for BNSs, $0.002<e_{\rm
r}<0.066$ for NSBHs, and $0.005<e_{\rm r}<0.15$ for BBHs. The reason
is that more GW energy is contributed from the low-frequency orbits
into the frequency band of aLIGO. Furthermore, our results show that
the GW-SGRB co-detection rate is nearly $\sim100$ per year for BBHs
(which could be $\sim 32$ times higher for $30 M_\odot$ of the black
hole mass), which is much higher than the recent observations of the
BBH merger rate (4.8/year for the first observing run of the aLIGO,
4.5/year for the second observing run). Therefore, even if
all the observed BBHs are SGRB progenitors (let alone only two BBHs
may be possible SGRB progenitors), only a small fraction of BBHs
could be SGRB progenitors, unless they have very large residual
eccentricities before merger (e.g., $e_r>0.7$ with BBH merger rate
of $\sim 4$/year). Recent observation of GW170817 confirmed the coalescing model for SGRBs, and the GW-SGRB
co-detection rate is about 1.5 per year (one event observed in 8
months for the second run of aLIGO), which may implies that not all
SGRBs are originated from the mergers of BNSs (0.8 event per year at
most), a fraction of BHNSs and BBHs may also be their origin (as
shown in figure \ref{figure 4}).
%For different SGRB progenitors, the  co-detection rates
%could be $\sim1.4$ times higher than the circular case, e.g., 120
%$\yr^{-1}$ for BBH , 113 $\yr^{-1}$ for SFR2, and 158 $\yr^{-1}$ for
%SFR3.

\section{Conclusion and Discussion}

Firstly, we derived the SGRB rate from the \emph{Swift}
observations, and the local SGRB rate is consistent with the
previous works for different SFR models \citep{Guetta09, Siellez14,
Clark15, Wanderman15}. Our results show that a constant time delay
distribution between the SGRB rate and SFR is preferred. The power
law time delay distribution is disfavored by the observations,
because it predicts too much dim SGRBs at high-redshifts. However,
we stress that the narrow constant time delay distribution could be
resulted from the lack of high-redshift SGRBs (e.g., $z>1.5$). More
high-redshift bursts is sufficient to make a difference.

Secondly, we studied the effect of eccentricity on SGWBs for
different types of SGRB progenitors, such as BNSs, NSBHs, and BBHs.
We found that SGWBs are greatly suppressed at low frequencies (e.g.,
$f\lesssim1 {\rm Hz}$) for eccentric orbit binaries.
Particularly, SGWBs from binaries with high residual
eccentricities (e.g., $e_{\rm r}\gtrsim0.1$ for BNSs) are hard to be
detected, because the radiated GW frequencies are above the
detection frequency of aLIGO/aVirgo. The chosen of different SFR
models have little effect on SGWBs, because the total SGRB number is
only determined by the observation. Anyway, it is possible that SGWB
from BBHs could be detected by aLIGO for one year of observation.
Moreover, SGWB is severely dependent on the merger rate of SGRB
progenitors and the black hole mass. For one point, the merger rate
could be 16 times lower for $\theta_j=20^{\circ}$ while comparing
with $\theta_j=5^{\circ}$ \citep[e.g., ][]{Chen13}, therefore, the
SGWB could be 16 times weaker, as shown in the right panel of figure
2. For another point, the SGWB could be $\sim6.2$ times higher if
BBH mass is assumed to be 30 $M_\odot$.

Finally, we calculated the GW-SGRB co-detection rates. On one hand,
we found that the co-detection rates are greatly reduced for
binaries with high residual eccentricities (e.g., $e_{\rm r}\gtrsim
0.2$). And we suggested that the extremely low GW-SGRB events may be
caused by the high residual eccentricities of binaries before
merger. On the other hand, the co-detection rates could be 1.4 times
higher than the circular case for binaries with residual
eccentricities within some particular ranges, shown as the bumps in
figure \ref{figure 4}. Additionally, we have observed two
possible and one confirmed GW-SGRB events: GW150914
\citep{Abbott16a, Connaughton16, Savchenko16}, GW170104
\citep{Verrecchia17}, and GW170817. The average event rate is 2.8
per year (three events in 13 months of the first and second run of
aLIGO). Comparing with our results in figure 4, it may implies that
not all SGRBs are originated from the mergers of BNSs, mergers of
BHNSs and BBHs may also be their origin. However, based on the
recent observations of BBH merger events and the GW-SGRB
co-detection events, our prediction of $\sim100$ GW-SGRB events per
year (32 times higher for $30 M_\odot$ of the BH mass) from BBH
merger seems too high. This means that only a small fraction
of BBHs could be SGRB progenitors, unless they have very large
residual eccentricities before merger (e.g., $e_r>0.7$ with BBH
merger rate of $\sim 4$/year). Specially, we should note that the
effect of the eccentricity on the co-detection rate is severely
dependent on the sensitivity of the GW detectors. The higher the
sensitivity of the GW detectors are, the more low eccentric orbit
binaries will be observed, and the plateau in figure \ref{figure 4}
will be shorten.

\section*{Acknowledgements}

We thank the anonymous referee for detailed comments and
suggestions. This work is supported by the National Basic Research
Program of China (973 Program, grant No. 2014CB845800) and the
National Natural Science Foundation of China (grants 11422325,
11373022, 11633001 and 11673008), the Excellent Youth Foundation of
Jiangsu Province (BK20140016), Jiangsu Planned Projects for
Postdoctoral Research Funds (0201003406).

\end{document}